\documentclass[prl,twocolumn,amssymb]{revtex4}
\usepackage{graphicx}
\usepackage{epstopdf}
\usepackage{dcolumn}
\usepackage{bm}
\usepackage{wasysym}

\newcommand{\beq}{\begin{equation}}
\newcommand{\eeq}{\end{equation}}
\newcommand{\beqn}{\begin{eqnarray}}
\newcommand{\eeqn}{\end{eqnarray}}

\begin{document}
\title{Majorana liquids: the complete fractionalization of the electron}
\author{Cenke Xu}
\affiliation{Department of Physics, Harvard University, Cambridge,
MA 02138, USA}

\author{Subir Sachdev}
\affiliation{Department of Physics, Harvard University, Cambridge,
MA 02138, USA}

\begin{abstract}
We describe ground states of correlated electron systems in which
the electron fractionalizes into separate quasiparticles which
carry its spin and its charge, and into real Majorana fermions
which carry its Fermi statistics. Such parent states provide a
unified theory of previously studied fractionalized states: their
descendants include insulating and conducting states with neutral
spin $S=1/2$ fermionic spinons, and states with spinless fermionic
charge carriers. We illustrate these ideas on the honeycomb
lattice, with field theories of such states and their phase
transitions.

\end{abstract}
 \maketitle

The study of two-dimensional quantum antiferromagnets has proved
to be fertile ground for finding many-electron states whose
excitations do not carry all the quantum numbers of the electron
\cite{pwa}. This phenomenon is often referred to as
``spin-charge'' separation. It leads to some of the most
non-trivial examples of quantum entanglement at long scales, and
is crucial for the understanding of a variety of correlated
electron materials, and for designing topological quantum
computers.

Upon fractionalizing the electron into its spin and charge, one is
faced with the decision of locating its Fermi statistics. In 
early theories of gapped spin liquid states of insulating,
frustrated, antiferromagnets,
distinct physical motivations led to two main pictures:\\
({\em i\/}) A picture of projecting out doubly-occupied and vacant
sites from a free electron Slater determinant produced an
attachment of Fermi statistics to spin, leading to \cite{wen1}
spin liquids with neutral, spin $S=1/2$ excitations
(`spinons') which are fermions. \\
({\em ii\/}) A picture of quantum-`disordering' magnetically
ordered states produced an attachment of Fermi statistics to charge,
leading to \cite{rstl} spin liquids with bosonic spinons.\\
This dichotomy persists when we consider insulating spin liquids
with gapless excitations, and also to metallic or superconducting
fractionalized states. Thus in the first picture we have
`algebraic spin liquids' (ASL) with Dirac points or Fermi surfaces
of neutral fermionic spinons
\cite{wenlee,herm2,hermele,florens,ssv}; while in the second
picture we have gapless bosonic spinons and/or gapless, spinless
fermionic charge carriers \cite{senthil,rkk2,sdwsu2},
 including `algebraic charge liquids' (ACL).

This paper will provide a unification of these seemingly divergent
pictures. We will argue that the states above descend from a
common parent `Majorana liquid', in which there is complete
fractionalization of the electron into components carrying its
spin, charge, and Fermi statistics. No choice is made in the parent liquid
regarding
attachment of Fermi statistics, which is instead carried by real
Majorana fermions which carry neither spin nor charge. While we
will introduce our approach generally, a specific
realization of the theory will be provided on the honeycomb
lattice. We will also relate our results
to a recent numerical study of the Hubbard model on the honeycomb
lattice \cite{meng}.

We motivate our approach by starting from the second picture
\cite{sdwsu2}. Here, we imagine there is some preferred local
magnetic order, and transform the electron ($c_\sigma$, $\sigma =
\uparrow, \downarrow$) into a rotating reference frame determined
by the orientation of the local magnetic order
\cite{shraiman}:
\begin{equation}
\left( \begin{array}{c} c_{\uparrow} \\ c_{\downarrow}
\end{array} \right) = \mathcal{R}_z \left( \begin{array}{c} f_{1}
\\ f_{2} \end{array} \right)~~;~~\mathcal{R}_z \equiv  \left(
\begin{array}{cc} z_{\uparrow} & -z_{\downarrow}^\ast \\
z_{\downarrow} & z_{\uparrow}^\ast \end{array} \right). \label{cz}
\end{equation}
Here the spinful bosons $z_\sigma$ define a SU(2) rotation matrix
$\mathcal{R}_z$ determined by the magnetic order, and $f_1 , f_2$
are spinless fermions which carry the charge of the electron. It
is also useful to note the various global and gauge symmetries
associated with Eq.~(\ref{cz}). The global spin rotation
SU(2)$_{\rm spin}$ acts as a left multiplication on
$\mathcal{R}_z$, while the global charge U(1)$_{\rm charge}$ is
carried by $f_{1,2}$. In addition there is a local SU(2)$_{s,g}$
gauge invariance: this acts as a right multiplication of
$\mathcal{R}_z$ and the doublet $(f_1, f_2)$ transforms as its
fundamental.

Now let us turn to the first picture \cite{wenlee,hermele}. We can
view this as a transformation into a rotating reference frame in
the pseudo-spin space of particle-hole transformations of the
electron \cite{affleck}. For the Hubbard model on bi-partite
lattices at half-filling, there is in fact a global SU(2)$_{\rm
pseudospin}$ which contains U(1)$_{\rm charge}$ as a subgroup;
however our approach only assumes U(1)$_{\rm charge}$ symmetry in
general. Paralleling Eq.~(\ref{cz}) we now have
\cite{wenlee,hermele}:
\begin{equation}
\left( \begin{array}{c} c_{\uparrow} \\ c_{\downarrow}^\dagger
\end{array} \right) = \mathcal{R}_b \left( \begin{array}{c} f_{1}
\\ f_{2}^\dagger \end{array} \right)~~;~~\mathcal{R}_b \equiv
\left( \begin{array}{cc} b_1^\ast & b_2^\ast \\ - b_2 & b_1
\end{array} \right). \label{cb}
\end{equation}
Independent of the existence of a global SU(2)$_{\rm pseudospin}$
symmetry, this parameterization introduces a  local SU(2)$_{c,g}$
gauge invariance: this acts as a right multiplication of
$\mathcal{R}_b$ and the doublet $(f_1 , f_2^\dagger )$ transforms
as its fundamental. This SU(2)$_{c,g}$ gauge invariance played a
crucial role in the classification of various spin liquid states
with fermionic spinons \cite{wen2,kitaev}.

There is a natural and simple unification of Eqs.~(\ref{cz}) and
(\ref{cb}). We write the complex fermions in terms of 2 sets of
real fermions $\zeta_a$ and $\chi_a$, with $a=1\ldots 4$, by
$c_\uparrow = \zeta_1 + i \zeta_2$ and $c_\downarrow = \zeta_3 + i
\zeta_4$ and similarly between $f_{1,2}$ and $\chi_a$. Then we
have
\begin{equation}
\zeta = \mathcal{R} \, \chi
\label{zeta}
\end{equation}
where $\mathcal{R}$ is a real SO(4) matrix. This shows that a
combination of Eqs.~(\ref{cz}) and (\ref{cb}) enjoys a SO(4)$_g$
gauge invariance; it will become evident below that ${\rm SO}(4)_g
\sim {\rm SU}(2)_{s,g} \otimes {\rm SU}(2)_{c,g}$. As before, the
SO(4)$_g$ gauge invariance acts as a right multiplication of
$\mathcal{R}$, $\chi$ is a fundamental of SO(4)$_g$, and global
symmetries act as left multiplication of $\mathcal{R}$. For an
explicit form of the global symmetries, let us represent the $4
\times 4$ real matrices as tensor products of 2 sets of $2 \times
2$ Pauli matrices: $s^{\alpha}$ acting on the $\uparrow$,
$\downarrow$ space ($\alpha = x,y,z$), and $\rho^{\alpha}$ acting
on the $\mbox{Re}[c]$, $\mbox{Im}[c]$ space. Then the global
SU(2)$_{\rm spin}$ symmetry is generated by
\begin{equation}
S^x = s^x \rho^y~~,~~S^y = s^y~~,~~S^z = s^z \rho^y .
\label{Ss}
\end{equation}
The global pseudo-spin transformations are generated by
\begin{equation}
T^x = s^y \rho^z~~,~~T^y = s^y \rho^x ~~,~~T^z = \rho^y ;
\label{Ts}
\end{equation}
here $T^z$ generates U(1)$_{\rm charge}$, while $T^{x,y}$ generate
the remaining pseudo-spin transformations which need not be a
symmetry of the Hamiltonian. The matrices in Eqs.~(\ref{Ss}) and
(\ref{Ts}) are the 6 generators of SO(4); note also that the
$S^{\alpha}$ commute with the $T^{\alpha}$, realizing the
factorization into ${\rm SU}(2) \otimes {\rm SU}(2)$. To complete
our formulation, we can express $\mathcal{R}$ in terms of the
complex bosons in $\mathcal{R}_z$ and $\mathcal{R}_b$. Writing
$z_\uparrow = \phi^s_0 - i \phi^s_3$, $z_\downarrow = - \phi^s_2 -
i \phi^s_1$, and $b_1 = \phi^c_0 + i \phi^c_3$, $b_2 = - i
\phi^c_2 + \phi^c_1$ where $\phi^c_a$ and $\phi^c_a$ are real
scalars, we have
\begin{eqnarray}
\mathcal{R} &=& Z_s \otimes Z_c \nonumber \\
Z_s &=& \phi^s_0 + i \phi^s_1 S^x + i \phi^s_2 S^y + i \phi^s_3 S^z \nonumber \\
Z_c &=& \phi^c_0 + i \phi^c_1 T^x + i \phi^c_2 T^y + i \phi^c_3 T^z . \label{Rz}
\end{eqnarray}

Eqs.~(\ref{zeta}) and (\ref{Rz}) contain our general statement of
electron fractionalization: the electron $\zeta$ decomposes into
the bosonic fields $Z_s$ and $Z_c$ which carry its global SU(2)$_{\rm spin}$
and U(1)$_{\rm charge}$ quantum numbers
respectively, and into the Majorana fermion $\chi$ carrying the
Fermi statistics. The resulting theory has a ${\rm SO}(4)_g = {\rm
SU}(2)_{s,g} \otimes {\rm SU}(2)_{c,g}$ gauge invariance: $Z_s$
and $\chi$ carry SU(2)$_{s,g}$ charges, and $Z_c$ and $\chi$ carry
SU(2)$_{c,g}$ charges. 

Different patterns of breaking the ${\rm SO}(4)_g$ gauge invariance and global symmetries
lead to a plethora of possible phases, and we present a broad classification:\\
(I) Phases with conventional excitations are obtained when
both $Z_c$ and $Z_s$ are condensed. In such phases, we can always choose a 
${\rm SO}(4)_g$ gauge
in which $Z_c=Z_s=1$, and then it becomes clear from Eq.~(\ref{zeta}) that 
the fermion $\chi$ has just the same quantum numbers as the electron $\zeta$.
By condensing various fermion bilinears (just as in conventional Hartree-Fock/BCS theory),
we can obtain 
Fermi liquids, semi-metals, antiferromagnets, valence bond solids
(VBS), superconductors, or
quantum spin Hall states \cite{kanemele,grover}.\\
(II) When $Z_s$ is condensed, we can use the SU(2)$_{s,g}$
gauge invariance to set $Z_s = 1$. Then Eq.~(\ref{zeta}) reduces to 
Eq.~(\ref{cb}), and we therefore reproduce the phases
of Refs.~\cite{wenlee,wen2}
on the square lattice, and those of Ref.~\cite{hermele} on the honeycomb lattice, 
including the ASLs. In these phases, the fixing of the SU(2)$_{s,g}$ gauge transfers
the global SU(2)$_{\rm spin}$ quantum numbers from $Z_s$ to $\chi$, while the 
U(1)$_{\rm charge}$ quantum numbers remain on $Z_c$. Thus these phases have
neutral fermionic spinons, and bosonic charge carriers.\\
(III) A complementary situation is realized when $Z_c$ is condensed.
Now we can set $Z_c=1$, Eq.~(\ref{zeta}) reduces to Eq.~(\ref{cz}),
global U(1)$_{\rm charge}$ quantum numbers are transferred from $Z_c$ to $\chi$, 
and so we obtain neutral bosonic spinons and fermionic charge carriers.
We reproduce phases of Refs.~\cite{rkk2,sdwsu2} on the square lattice, and 
Ref.~\cite{fawang2010} on the honeycomb lattice, 
including the ACLs. \\
(IV) When neither $Z_c$ and $Z_s$ are condensed, we
can obtain phases in which the $Z_c$, $Z_s$, and $\chi$ are 
separate elementary excitations, carrying the charge, spin,
and Fermi statistics of the electron respectively. These are the
Majorana liquids of this paper. These elementary excitations all carry
gauge quantum numbers, and so the stability of such phases requires that
gauge forces not be confining: we will describe specific examples of deconfinement
mechanisms below.

The remainder of the paper applies our general theory to the
half-filled, extended Hubbard model on the honeycomb lattice. For
weak interactions, we have a conventional semi-metal with
electronic excitations at 2 Dirac points in the Brillouin zone:
this is as realized in graphene. For strong interactions, there is
convincing numerical evidence \cite{meng} for an insulator with
collinear, two-sublattice antiferromagnetism (N\'eel order). The
simplest possibility is that there is a direct transition between
the these two category I phases \cite{herbut}. However, recent
numerical studies \cite{meng} indicate there may be an
intermediate phase.

We begin our analysis by describing the free electron spectrum in
the semi-metal phase, associated with the Hamiltonian
\begin{equation}
H_0 = - t \sum_{\langle ij \rangle, } c^\dagger_{i \sigma} c_{j \sigma} \label{H0}
\end{equation}
where $i$ are sites on the honeycomb lattice, and $\langle ij
\rangle$ refers to nearest neighbors. We take the low energy limit
of $H_0$ in the standard manner, obtaining two valleys of
two-component Dirac fermions. Explicitly, we expand the electron
at two Dirac valleys by $d_{1,2} = e^{i \vec{Q}_{1,2} \cdot
\vec{r}} c$ (where $\vec{Q}_{1,2} = \pm (\frac{4\pi}{3}, 0) $ are
the wavevectors of the valleys), and introduce Pauli matrices
$\tau^{\alpha}$ and $\mu^{\alpha}$ which act on the sublattice and
valley spaces respectively. Then, after introducing real Majorana
fermions $\zeta_a$ as the real and imaginary parts of
$e^{i\frac{\pi}{4}\tau^x}e^{i\frac{\pi}{4}\mu^x} (d_1, i\tau^y
d_2)^t$, we obtain the continuum Lagrangian
\begin{equation}
\mathcal{L}_0 = \sum_{a = 1}^8 \bar{\zeta}_a \gamma_\mu\partial_\mu \zeta_a . \label{L0}
\end{equation}
Here $\mu$ is a 2+1 dimensional spacetime index, and the Dirac
$\gamma$ matrices are $(\gamma_0, \gamma_1, \gamma_2) =
\tau^y,\tau^z, \tau^x$. In contrast to the single-site Majorana
fermion in Eq.~(\ref{zeta}), here the fermion field $\zeta$ has
additional components associated with the sublattice and valley
spaces. The sublattice index is equivalent to the spacetime Dirac
index and has been left implicit, and from now on the `flavor'
index $a=1 \ldots 8$ accounts for the spin, pseudospin, and valley
indices; thus for each $a$, $\zeta_a$ is now a 2-component
Majorana spinor. We can now decompose $\zeta$ as in
Eq.~(\ref{zeta}), with $\chi_a$ having the same spinor structure
as $\zeta_a$, while $\mathcal{R}$ is as in Eq.~(\ref{Rz}).

We obtain our parent {\em algebraic Majorana liquid\/} (AML) when 
interactions beyond those in $H_0$ leave both $Z_s$
and $Z_c$ un-condensed and realized as gapped quanta which
carry spin and charge respectively. The AML is in our category IV above,
and at energies below the spin and charge gaps, it has 
gapless, relativistic, Majorana fermions $\zeta_a$ coupled to
emergent SU(2) gauge fields $A^\alpha_{s,\mu}$ and
$A^\alpha_{c,\mu}$ associated with the  ${\rm SU}(2)_{s,g} \otimes
{\rm SU}(2)_{c,g}$ gauge invariance:
\begin{equation}
\mathcal{L}_{\rm AML} = \bar{\chi} \gamma_\mu \left( \partial_\mu
- i A^\alpha_{s,\mu} S^\alpha - i A^\alpha_{c,\mu} T^\alpha
\right) \chi \label{aml}
\end{equation}
The stability of the AML requires that the gapless $\chi$ fermions
suppress monopoles, and so that the ${\rm SU}(2)_{s,g} \otimes
{\rm SU}(2)_{c,g}$ gauge forces are not confining \cite{herm2,ran}.
Such monopole suppression happens for a sufficiently large number of
gapless fermion flavors, and it is not known if the 8 real fermion flavors here
are sufficient. Assuming deconfinement, the AML
has a gap to all spin and charge excitations, and has gapless
Majorana fermions which carry only energy.

Whether or not the AML is stable, we can 
use it to describe a very large number of descendant phases. The rest
of the paper will note some interesting or physically relevant
examples. 

First, we can expect that the ${\rm SU}(2)_{s,g} \otimes
{\rm SU}(2)_{c,g}$ gauge forces lead to the analog of chiral
symmetry breaking, and the simplest possibility is the appearance
of a O(8) invariant $\bar{\chi} \chi$ condensate. Such a
condensate leads to a fermion mass gap, and breaks time-reversal
symmetry, leading to a {\em chiral Majorana liquid\/}, also in category IV. 
The fermions generate Chern-Simons terms for the gauge fields,
and this leads to deconfinement \cite{csl} for the gapped $Z_c$, $Z_s$, and $\chi$
excitations.
There is
non-zero spin chirality $\vec{S}_1 \cdot(\vec{S}_2 \times
\vec{S}_3) + \cdots$, and also nonzero electrical currents on the
lattice. However, the physical current $\bar{\zeta}\zeta$ is
suppressed from $\bar{\chi}\chi$: \beqn \langle \bar{\zeta}\zeta
\rangle \sim \langle \bar{\chi}\chi \rangle \langle Z^\dagger_{s,
i}Z^\dagger_{c, i} Z_{s, j}Z_{c, j}\rangle_{\ll i,j \gg}, \eeqn
$i$ and $j$ are two next nearest neighbor sites, therefore the
expectation value $\langle Z^\dagger_{s, i}Z^\dagger_{c, i} Z_{s,
j}Z_{c, j}\rangle$ is expected to be small when $Z_s$ and $Z_c$
are both gapped.

There are also a large number of possible Higgs phases. One interesting
example is the Higgs condensate of the vector $H^\alpha$:
\begin{equation}
H^{\alpha} = \bar{\chi}\  \rho^y S^\alpha \  \chi. \label{H}
\end{equation}
If $H^\alpha$ had involved bilinears of the original electron
$\zeta$, its condensate would break spin rotation invariance and
lead to a quantum spin Hall phase \cite{grover}. In the present
situation, the $H^\alpha$ in Eq.~(\ref{H}) does not carry any
global quantum numbers. and its Higgs condensate does not break
any global symmetries. The resulting phase is in fact a {\em $Z_2$
Majorana liquid\/}, and is in category IV, as we now argue. The $H^\alpha$ condensate
breaks the $\mathrm{SU}(2)_{s,g} \otimes \mathrm{SU}(2)_{c,g}$
gauge invariance to $\mathrm{U}(1)_{s,g} \otimes
\mathrm{U}(1)_{c,g}$; if we choose $H^\alpha \propto (0,0,1)$,
then the U(1)'s are generated by $S^z$ and $T^z$. Thus the low
energy theory of this phase is
\begin{equation}
\mathcal{L}_{Z_2} = \bar{\chi} \gamma_\mu \left( \partial_\mu - i
A^z_{s,\mu} S^z  - i A^z_{c,\mu} T^z \right) \chi + m\, \bar{\chi}
\rho^y S^z \chi , \label{Z2}
\end{equation}
where the fermion mass term is induced by the $H^\alpha$
condensate. We can further integrate out the massive $\chi$
fermions, and then using the analog of the arguments in
Ref.~\cite{grover}, we find that the physics is controlled by a
mutual Chern-Simons term: \beqn \mathcal{L}_{cs} =
\frac{2i}{2\pi}\epsilon_{\mu\nu\rho} A^z_{c,\mu}\partial_\nu
A^z_{s,\rho}. \label{lcs} \eeqn 
As discussed elsewhere \cite{xs}, with such a term, the gauge
forces are quenched, and the matter fields carry only $Z_2$ gauge charges.
The $Z_2$ gauge field endows mutual
statistics between excitations with $S^z$ and $T^z$ charges, for
instance between the spin- and charge-carrying bosons, $Z_s$ and $Z_c$, as in Ref.~\cite{kouqiweng}. 

The last two category IV phases above, the chiral and $Z_2$ Majorana liquids, 
are attractive candidates for the intermediate state in Ref.~\cite{meng} between the semi-metal
and the N\'eel insulator.  
They have gapped $Z_c$, $Z_s$, and $\chi$ excitations, and we
have demonstrated that the charge, spin, and Fermi statistics they 
carry remain deconfined. These phases treat the charge and spin
excitations at an equal footing (unlike the proposal in Ref.~\cite{fawang2010}), 
and so they are appropriate for the 
vicinity of the metal-insulator transition. The chiral Majorana liquid
has weak spontaneous spin chirality and electrical currents, which have
not been detected so far. The $Z_2$ Majorana liquid has no broken symmetry,
and so remains compatible with all existing computations.

Let us now turn to category II. As noted earlier, 
such phases have $\langle Z_s \rangle \neq 0$, and on the honeycomb
lattice our theory reduces to that of Hermele \cite{hermele}.
He found an SU(2)$_{c,g}$ ASL insulator: at low energies, this is described by the theory of the AML
in Eq.~(\ref{aml}), but with the SU(2)$_{s,g}$ gauge fields $A^\alpha_{s,\mu} = 0$
because of the $Z_s$ condensate. As with the AML, this is stable only for 
sufficiently large number of fermion flavors. The $\chi$ fermions now carry spin
(explained in (II)),
and so the spin excitations are gapless.
The
intermediate phase of Ref.~\cite{meng} has a spin gap in
addition to the charge gap, and this is not compatible with an ASL.

Finally, we turn to category III, where we have $\langle Z_c \rangle \neq 0$. 
These were discussed in Ref.~\cite{sdwsu2} for the square lattice,
and we generalize the discussion here to the honeycomb lattice.
As in category II above, we begin with the AML in Eq.~(\ref{aml}), and now derive
a SU(2)$_{s,g}$ ACL obtained by setting the SU(2)$_{c,g}$ gauge fields $A^\alpha_{c,\mu} = 0$.
This phase has a spin gap, but the fermions $\chi$ now form gapless
excitations which carry charge (as explained in (III)), again incompatible with the phase of Ref.~\cite{meng}.
 As for the AML, the ACL can lead to a `chiral charge liquid' with a
$\bar{\chi} \chi$ condensate.

\begin{figure}[t]
\begin{center}
\includegraphics[width=2.4in]{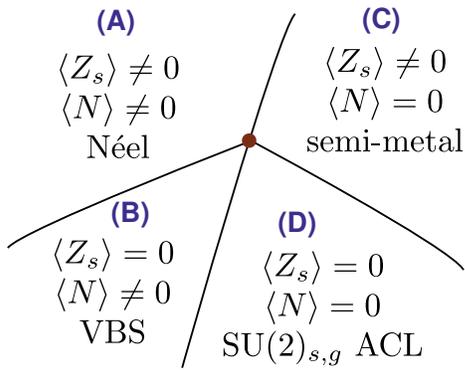}
\caption{Some phases of the half-filled honeycomb lattice.
All phases above have the 
condensate $\langle Z_c \rangle \neq 0$, and so can be described by Eq.~(\ref{cz}). Phase D is 
category III and is described by
Eq.~(\ref{aml}) but with $A_{c,\mu}^\alpha = 0$: it has a spin gap, and
gapless, spinless critical fermions at the Dirac points carrying
the electronic charge. Phases A and C are conventional and so in category I:
C has gapless electron-like excitations
at the 2 Dirac points, which are gapped in A. Phase B is described
by Eq.~(\ref{B}), which undergoes monopole-induced confinement to VBS order.}
\label{figacl}
\end{center}
\end{figure}

Other phases which descend from the SU(2)$_{s,g}$ ACL are
shown in Fig.~\ref{figacl}.
The simplest of these is the semi-metal phase C: this is the category I phase with both 
$Z_c$ and $Z_s$ condensates, as noted earlier. For the remaining
phases in Fig.~\ref{figacl} we have to consider a Higgs condensate of the field
\begin{equation}
N^{\alpha} = \bar{\chi}\  \mu^y S^\alpha \  \chi. \label{N}
\end{equation}
This choice is motivated by the fact that if we replace $\chi$ by the electron $\zeta$,
then Eq.~(\ref{N}) is the conventional N\'eel order;
this replacement is permitted when both $Z_c$ and $Z_s$ are condensed.
This is the case in phase A, which is then a category I phase with N\'eel order.
The transition between phases C and A involves the order
parameter and the gapless Dirac electrons, but no gauge fields; it
is in the class of field theories studied in Ref.~\cite{herbut}.

Now we consider the remaining phase B in Fig.~\ref{figacl}. By a gauge
transformation, we orient the $N^\alpha $ condensate along
$(0,0,1)$; such a condensate breaks the SU(2)$_{s,g}$ gauge
invariance down to U(1)$_{s,g}$, leaving only the $A^z_{s,\mu}$
gauge field active. The neutral $Z_s$ spinons are gapped, and so
the low energy theory of phase B (following Eq.~(\ref{Z2})) is
\begin{equation}
\mathcal{L}_{\rm B} = \bar{\chi} \gamma_\mu \left( \partial_\mu -
i A^z_{s,\mu} S^z \right) \chi + m \, \bar{\chi} \mu^y S^z \chi .
\label{B}
\end{equation}
As after Eq.~(\ref{Z2}), we can integrate out the massive
fermions, but now find only a Maxwell term for the $A^z_{s,\mu}$
gauge field. Thus the low energy theory of phase B has a gapless,
relativistic U(1)$_{s,g}$ photon. In the absence of gapless matter,
it is known that the monopoles in such a gauge field condense, and lead to long-range order
determined by the quantum numbers of the monopole \cite{sachdev1990}.
We will describe the computation of monopole quantum numbers elsewhere, showing that a kekul\'e 
type valence bond solid (VBS) order develops. The same conclusion is
reached by approaching phase B from phase A \cite{sachdev1990}. Thus,
while phase B started out in category III, it ultimately becomes category I VBS state.

This paper has unified two previously divergent approaches to the
study of the electron fractionalization: those with fermionic
\cite{wen1,wenlee,herm2,hermele,florens,ssv,wen2,kitaev} versus
bosonic \cite{senthil,rkk2,sdwsu2} spinons. We have applied the
theory to the honeycomb lattice and predicted new phases of
possible relevance to recent numerical results \cite{meng}.

This research was supported by the National Science Foundation
under grant DMR-0757145, by the FQXi foundation, and by a MURI
grant from AFOSR.

\end{document}